\documentclass[11pt,twoside]{article}

\usepackage{asp2006}
\usepackage{epsf}
\usepackage{psfig}
\usepackage{lscape}
\usepackage{graphicx,picins}

\begin{document}

\title{Interactions between a massive planet and a disc}   

%
%
%

\author{Pau Amaro-Seoane\altaffilmark{1,2}, Ignasi Ribas\altaffilmark{2}, Ulf L{\"o}ckmann\altaffilmark{3}, Holger Baumgardt\altaffilmark{3}}
\altaffiltext{1}{Corresponding author, Pau.Amaro-Seoane@aei.mpg.de, Max-Planck Institut f{\"u}r Gravitationsphysik (Albert Einstein-Institut), Am M{\"u}hlenberg 1, D-14476 Potsdam}
\altaffiltext{2}{Institut de Ci{\`e}ncies de l'Espai, Campus UAB, Torre C-5, parells, 2na planta, ES-08193 Bellaterra, Barcelona}
\altaffiltext{3}{Argelander-Institut f{\"u}r Astronomie, Auf dem H{\"u}gel 71 D - 53121 Bonn}

\begin{abstract} 
We analyse the potential migration of massive planets forming far away from an
inner planetary system. For this, we follow the dynamical evolution of the
orbital elements of a massive planet undergoing a dissipative process with a
gas disc centred around the central sun.  We use a new method for
post-Newtonian, high-precision integration of planetary systems containing a
central sun by splitting the forces on a particle between a dominant central
force and additional perturbations. In this treatment, which allows us to
integrate with a very high-accuracy close encounters, all gravitational forces
are integrated directly, without resorting to any simplifying approach.
After traversing the disc a number of times, the planet is finally trapped into
the disc with a non-negligible eccentricity
\end{abstract}



\section{Introduction}
It has been suggested that the most massive planets in a planetary system can
be formed by a process of gas collapse, independently of metallicity, whilst
the lighter components would have formed via core accretion. This can lead to a
situation in which massive planets that have originated relatively far away
from the inner system, migrate inwards into it \citep[see][and contribution by
Font-Ribera et al. in this same volume]{Font-RiberaEtAl09}.  In this process,
these more massive planets with larger semi-major axis, cross the gas disc
centred around the central sun. When going through this dissipative process,
the planets lose kinetic energy because of the friction with the gas.  As a
consequence, the inner disc is heated up and the semi-major axis of the planet
shrinks. After some passages, the planet is trapped in the disc with a residual
eccentricity. We propose this scenario as a plausible way of explaining the
existence of massive planets distributed around a sun with non-zero
eccentricities. We give results based on high-accurate dynamical simulations
about the distribution of the orbital elements of the trapped objects in the
disc.  We find that the massive planets are typically captured after some $\sim
10^{5}$ yrs and the final eccentricity is non-negligible ($e \sim 0.1$).

\section{Method, results and conclusions}

We have recently developed an integrator {\tt bhint} specialised for dynamical
processes in the vicinity of a very massive particle, which relies in the
assumption that the very massive particle dominates the motion of the smaller
ones \citep{LoeckmannBaumgardt08}. For this new mechanism, we retain the
Hermite scheme as a basis. The bottleneck is, of course, the number of
particles to be used. Nonetheless, we resort to special-purpose hardware, the
GRAPE, a card specially developed to integrate the calculation of Newtonian
gravitational forces.  The peak performance of one of these cards is of 130
Gflop, roughly equivalent to 100 single PCs, which makes possible long
simulations with a realistic particle number.  

\parpic[r][r]{\mbox{\resizebox{0.45\hsize}{!}
        {\includegraphics{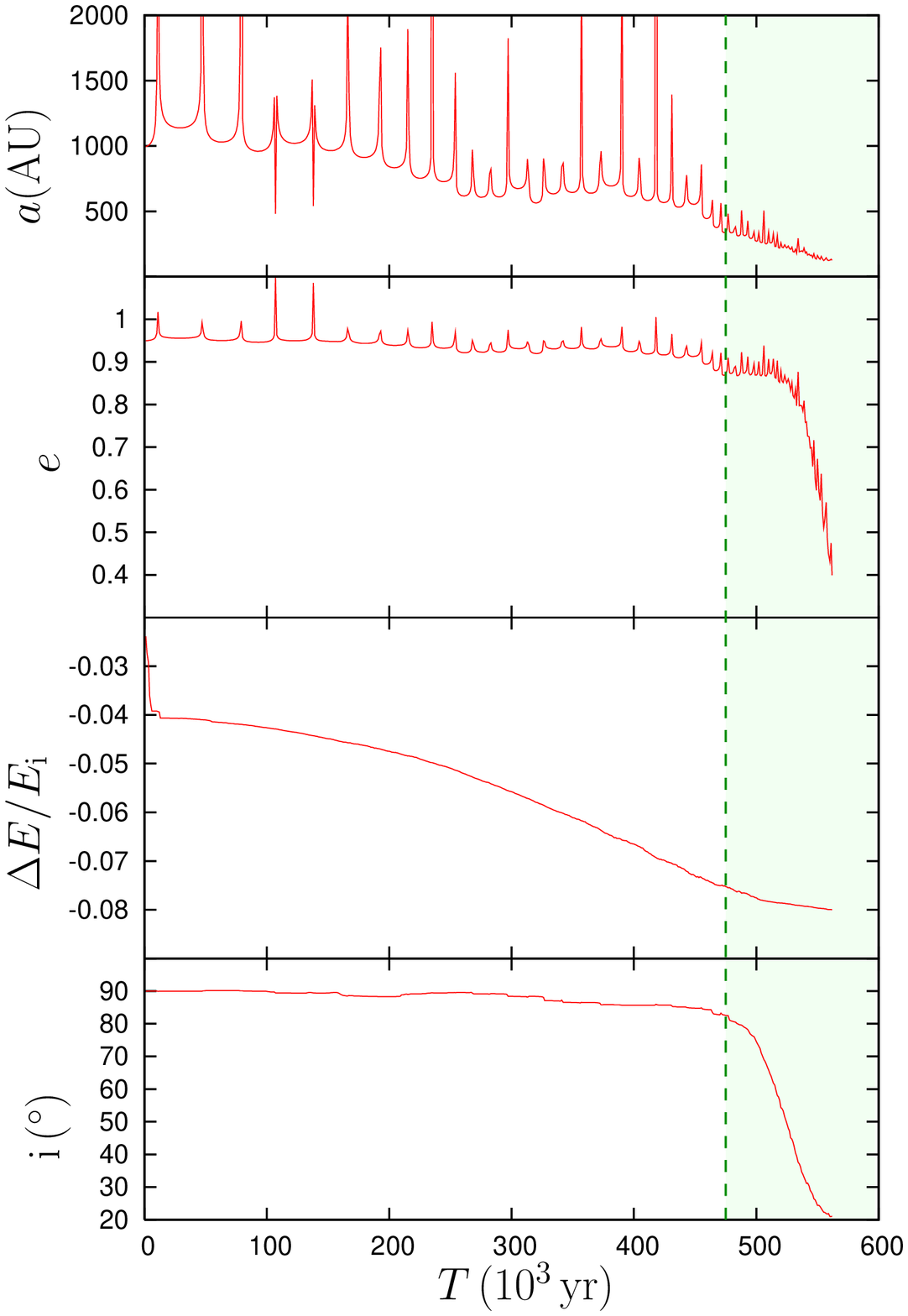}}}}

We set initially a disc made out of $10^3$ small particles which is ``hosting''
a sun in the centre and follows a simple $1/r$ profile. The integrated mass is
of some 5 Jupiters and the radius of some $30$ AU. The thickness of the disc is
of about the diameter of the central sun and has a gap around the central sun
which extends $0.1$ AU.  The mass of the central sun is $1\,M_{\odot}$. The
mass forming the disc are all single-mass. The massive planet, a massive
particle of 5 Jupiters is set in an orbit such that the initial eccentricity
with the sun is of $e = 0.95$.  Initially, the particle is 100 AU away from the
sun and the inclination angle is $i = 90$ degrees.  The system (disc plus
interloper) is integrated until the interloper is trapped by the disc.
In the figure we show the evolution of the orbital parameters. Whilst we cannot
discuss them in detail because of the publication limits, we note that after
some $4 \times 10^5$ yrs the inclination has almost not changed as compared to
its initial value. Then, after a short time of $75 \times 10^3$ yrs elapses, it
abruptly decays from almost 80 degrees to a very small number, to be finally
trapped in the disc after $5 \times 10^5$ yrs. The energy, whilst it decays
from the initial high value of 0.95, is of $e \sim 0.1$ when the massive planet
is totally trapped in the disc, within a final semi-major axis which is well
within the range of expectation.  An extended and detailed scrutiny of the
parameter space of this capture process we propose will be soon published
elsewhere \citep{ASEtAl09}.  The adressing of this scenario has direct bearing on our
understanding of planetary dynamics and migration mechanisms.

\end{document}